\documentclass{elsart}
\usepackage{amsmath,amssymb,graphics,epsfig}

\newcommand{\bra}[1]{\left\langle #1\right |}
\newcommand{\ket}[1]{\left| #1\right \rangle}
\newcommand{\braket}[2]{\left\langle #1 | #2 \right\rangle}

\begin{document}
\begin{frontmatter}
\title{Topological tunnelling with Dynamical overlap fermions}
\author[a]{Nigel Cundy}
\author[c]{Stefan Krieg,$^{\text{b},}$}
\author[c]{Thomas Lippert,$^{\text{b},}$}
\author[a]{and Andreas Sch\"afer}
\address[a]{Institut f\"ur Theoretische Physik, Universit\"at Regensburg, D-93040 Regensburg, Germany}
\address[b] {Department of Physics, University of Wuppertal,
  Germany.}
\address[c]{J\"ulich Supercomputing Centre,
    J\"ulich Research
 Centre, 52425 J\"ulich, Germany.}

\begin{abstract}
Tunnelling between different topological sectors with dynamical chiral 
fermions is difficult because of a poor mass scaling of the pseudo-fermion estimate of 
the determinant. For small fermion masses it is virtually 
impossible using standard methods. However, by projecting out the small Wilson eigenvectors from 
the overlap operator, and treating the correction determinant exactly, we 
can significantly increase the rate of topological sector tunnelling and 
reduce substantially the auto-correlation time. We present and compare a 
number of different approaches, and advocate a method which allows 
topological tunnelling even at low mass with little addition to the 
computational cost.
\end{abstract}
\end{frontmatter}
\section{Introduction}\label{sec:1}
In recent years the statistical precision of lattice simulations has reached 
such a high level that the complete control of systematic uncertainties 
became an issue of central importance. In this respect
the overlap fermion action\cite{Narayanan:1993ss,Neuberger:1998fp}
would be the  preferred choice, were it not for its massive 
computational needs.  
Unlike other formulations of lattice QCD it has an exact lattice chiral 
symmetry~\cite{Luscher:1998du} and index theorem
such that the combined continuum and chiral limits should not hold any 
unpleasant surprises. 
Pioneering investigations~\cite{Fodor:2003bh,Cundy:2005pi,DeGrand:2004nq,Hashimoto:2006rb} showed, however, that such simulations are 
faced with quite formidable challenges. While the rapid progress of
massively parallel computers will alleviate the problems~\cite{Fukaya:2007fb} 
the feasibility of really controlled simulations requires some 
algorithmic break-through.
In this paper we shall focus on one of the main algorithmic issues, 
of considerable concern in the literature and present a possible solution.

The overlap operator is defined as
\begin{gather}
D = (1+\mu) + (1-\mu)\gamma_5\epsilon(Q).
\end{gather}
$Q$ is constructed from any doubler-free Dirac operator $D_W$ which fulfils 
$\gamma_5$-hermiticity according to 
$Q=\gamma_5D_W(-\rho)$. For the Wilson Dirac operator this gives 
\begin{gather}
Q_{xy} = \gamma_5 \left[\delta_{xy}-\kappa\left( (1-\gamma_{\mu})U_{\mu}(x)
\delta_{y,x+\mu} + (1+\gamma_{\mu})U^{\dagger}_{\mu}(x-\mu)
\delta_{y,x-\mu}\right)\right].
\end{gather}
$\rho$ and $\mu$ are mass parameters related to the bare fermion mass 
by $m_{\rm bare}=2\mu\rho$. (Note that different definitions 
of $\rho$ are used in the literature.) 
In this work, we will always use the Wilson Dirac operator as kernel with 
$\rho = 1.5; \kappa = 1/(8-2\rho) = 0.2$.

The matrix sign function is defined as
\begin{gather}
\epsilon(Q) = \sum_{i}|\psi_i\rangle\langle \psi_i|\text{sign}(\lambda_i),
\end{gather}
where $|\psi_i\rangle$ and $\lambda_i$ are the eigenvectors and eigenvalues of $Q$ respectively, and 
the sum is over the complete set of eigenvectors. In practice, given that the calculation of the entire 
eigenvalue spectrum is impractical, one combines an approximation to the sign function, such as the 
Zolotarev rational approximation~\cite{vandenEshof:2002ms} for the bulk of the eigenvalue spectrum, 
with the spectral decomposition for the eigenvalues closest to zero, where no approximation can 
(realistically) be accurate enough without immense computational cost.
 
In terms of the Hermitian overlap operator $H = \gamma_5 D$,  and the gauge action $S_g[U]$ for a gauge field $U$, the lattice QCD partition function for two degenerate fermion flavours is
\begin{gather}
Z = \int d U \det (H^2[U,\mu])e^{-S_g[U]} = \int dU d\phi d\phi^{\dagger}e^{-S_g[U]-\phi^{\dagger}H^{-2}\phi},
\end{gather}
where we have used pseudo-fermion fields $\phi$ to approximate the determinant. The standard HMC 
algorithm~\cite{HMC} generates a new gauge field by introducing a momentum $\Pi$, updating the 
momentum and gauge field along the classical trajectory using a numerical integration algorithm 
(molecular dynamics), and finishing with a Metropolis step to ensure that the update of the gauge field 
satisfies detailed balance. The numerical integration must be reversible and ergodic, and approximately 
conserve the action minus the logarithm of the Jacobian of the integration. 

The QCD vacuum contains several homotopy classes, each distinguished by a Pontryagin index, the winding number or topological charge. In the continuum theory, this topological charge is an integer given by
\begin{gather}
Q_c = \frac{1}{32\pi^2}\int d^4x \tilde{F}_{\mu\nu} F_{\mu\nu}.
\end{gather}
On the lattice, discretization errors break the barriers between the topological sectors, and the 
topological charge is no longer well defined. However, it is possible to define a modified topological 
index in a number of different ways, each leading to  exactly  or approximately  integer values. 
These topological indexes can be defined either using the gauge fields or fermionic operators.
In all cases some means must be found to filter out ultra violet fluctuations
which can lead to a different assignment of topological charge for the same configuration.  
As the lattice spacing shrinks, these artefacts disappear, and all the topological indexes 
reduce to the topological charge in the continuum. Here, we shall use solely the overlap definition 
of the topological index,
\begin{gather}
Q_f = -\frac{1}{2}\text{sign} Q.\label{topologicalcharge}
\end{gather}

Solutions for gauge field vacua with various topological charges are well known, for example 
instantons~\cite{Instantons1,Instantons2}, calorons~\cite{Calorons}, vortexes~\cite{Vortices}, membranes~\cite{Monopoles1,Monopoles2} etc. For the 
sake of simplicity, we shall assume for the following discussion that the topological properties of 
the vacuum are described by instantons 
and anti-instantons, although the physical reality seems to be far 
more complex. The specific model adopted is irrelevant for the arguments 
given below.

To ensure ergodicity, a lattice simulation must sample all topological sectors in an unbiased manner. 
In fact, it must also sample all parts of a topological sector (which might or might not be 
connected) in an unbiased manner. Otherwise one introduces hardly controllable systematic 
uncertainties. To stress this fact let us remind that many aspects of 
low energy physics are intimately connected to topologically non-trivial field configurations. 
Examples are, e.g., the density of low lying Dirac eigenvalues, and thus the chiral condensate, 
and through the latter the pion mass. 
To sample all topological configurations in an unbiased manner, 
the Monte-Carlo method used must allow for
the creation and destruction of instanton/anti-instanton pairs. 
Having instantons fade in and out of the resolution of the topological charge operator is 
{\em not} enough. As the production and annihilation of instantons and anti-instanton 
is usually associated with eigenvalues wandering through the Ginsparg-Wilson circle towards 
the origin, or with a crossing from positive to negative eigenvalues,
using methods to suppress very small kernel eigenvalues~\cite{Kaneko:2006pa,Vranas:1999rz} results in a number of systematic uncertainties. Therefore, our strategy is to treat overlap fermions exactly, 
despite the large numerical cost. Furthermore, if, in a 
HMC simulation with overlap fermions, kernel eigenvectors with near-zero eigenvalues are 
artificially suppressed, it is very hard to measure the rate of creation and annihilation of topological objects
and thus the auto-correlation time. We aim to perform simulations with 
frequent topological index changes.

A topological index change, using the overlap definition, occurs, e.g., when one of the eigenvalues of 
the kernel operator $Q$ changes sign, as is clear from the definition in equation (\ref{topologicalcharge}). Not 
surprisingly, this leads to a discontinuity in the fermion operator, and thus a $\delta-$function in 
the fermionic force leading to a change of the kinetic energy by $-\Delta S$.


The standard method to incorporate this Dirac $\delta-$function force into the HMC is the transmission/reflection algorithm first published in ~\cite{Fodor:2003bh} and improved by our own work in ~\cite{Cundy:2005pi} and ~\cite{Cundy:2005mr} (we improved the energy conservation, and increased the transmission rate). For simplicity we shall here discuss a specific two dimensional algorithm taken from ~\cite{Cundy:2005pi}, although our final result, in a sufficiently large volume, is unaffected by the choice of algorithm (not withstanding the improvements of ~\cite{Cundy:2005mr}). In this particular model, the momentum is changed parallel to two orthonormal basis vectors $\eta_1$ and $\eta_2$. The energy conservation equation reads
\begin{gather}
\frac{1}{2}\Pi^2_+ = \frac{1}{2}\Pi_-^2 -\Delta S,
\end{gather}
and we satisfy it by using the update (with $(A,B)$ denoting the inner product of the fields $A$ and $B$) 
\begin{gather}
\Pi_+ = \Pi_- + \big(\eta_1(\eta_1,\Pi_-)+ \eta_2(\eta_2,\Pi_-)\big) \left(\sqrt{1-\frac{2\Delta S}{(\eta_1,\Pi_-)^2+ (\eta_2,\Pi_-)^2}}-1\right).
\end{gather}
Proof that this algorithm is area conserving and reversible with a suitable choice of $\eta_1$ and $\eta_2$ can be found in ~\cite{Cundy:2005pi}. When the term in the square root is positive, we can continue, and the topological charge is changed. When it is negative, we have to reflect off the potential wall, and there is no topological charge change. Assuming that the momentum is given by a Gaussian distribution, the transmission rate, the number of transmission steps divided by the total number of transmission and reflection steps is thus for a given $\Delta S$
\begin{align}
R =& \frac{\int d(\eta_1,\Pi)d(\eta_2,\Pi)\theta((\eta_1,\Pi)^2+ (\eta_2,\Pi)^2 -2\Delta S)e^{-(\eta_1,\Pi)^2/2-(\eta_2,\Pi)^2/2}}{\int d(\eta_1,\Pi)d(\eta_2,\Pi) e^{-(\eta_1,\Pi)^2/2-(\eta_2,\Pi)^2/2}} \nonumber\\=& \text{min}(1,e^{-\Delta S}).
\end{align} 
For a sufficiently large volume, the transmission rate is given by this formula for all possible energy and area conserving transmission algorithms. While it is possible to improve the transmission rate using non-area conserving algorithms~\cite{Cundy:2005mr}, the dependence on the mass remains the same. To have a large transmission rate, and thus small auto-correlation, it is necessary to reduce the action jump, $\Delta S$.

In section \ref{sec:1b}, we demonstrate that $\Delta S$ scales as O($\mu^{-2}$) when the fermion determinant is estimated by a single pseudo-fermion. Therefore at low masses, topological charge changes are practically impossible. While using multiple pseudo-fermions reduces this problem, it does not eliminate it.

However, it was noted in ~\cite{DeGrand:2004nq,Egri:2005cx} that the change in the actual fermion determinant, as opposed to the pseudo-fermion estimate of it, scales much better with the quark mass, and in section \ref{sec:1b} we will argue that the dominant terms in the mean discontinuity are O($1$). Therefore, if we were to use the actual determinant, rather than the pseudo-fermion estimate, the bad scaling of the topological auto-correlation time with the mass would not occur. Of course, in practice, this is impossible. 

However, there is another possibility, which is to factorise the determinant into two terms: a larger continuous determinant that can be treated with pseudo-fermions; and a smaller determinant that contains all the discontinuities and which can be treated exactly. In principle, this should combine the low $\Delta S$ of the exact determinant, with the better volume scaling of the pseudo-fermion estimate. Over the past year and a half, we have explored many different factorisation schemes, some inexact (in the sense that the pseudo-fermion term was still discontinuous, though the action jump was reduced) and quick; others exact but slower. Here, we will concentrate on one of these algorithms, which we call algorithm C. We presented preliminary results for algorithm C and a second approach, algorithm F, in ~\cite{Cundy:2007dp}.
We note that Stefan Schaefer independently proposed a similar method in ~\cite{Schaefer:2006bk}, though later abandoned it~\cite{Degrandschaeferpersonalcommunication}, partly in view of the difficulties which we shall discuss in section \ref{sec:3b}. We have tested our implementation of this method on $8^316$ lattices and $12^348$ lattices (although here we only present results from the $8^316$ lattices), and, using various tricks, have not encountered any difficulties. There might be problems as we move to larger lattice volumes, but we are currently working to avoid them.  

One obvious criticism of our approach is that we are focusing on increasing the transmission rate, while the auto-correlation time depends on the rate of topological charge change, which is the transmission rate multiplied by the number of attempted topological charge changes. However, we believe that the transmission rate is the key quantity when determining the efficiency of the algorithm. It takes time to perform the transmission/reflection step: we need to find the gauge field where the kernel operator has a zero eigenvalue; the kernel eigenvalues need to be recalculated; and several overlap inversions are needed to calculate $\Delta S$ and the fermionic forces, both for the transmission and reflection steps. Each transmission step costs a certain amount of time, but results in a change in the topological charge, and thus the cost is justified because it reduces the auto-correlation. When there is no transmission/reflection step, there is no topological charge change, but there is also no additional cost. For a reflection step, however, we both spend the computer time and get no tunnelling event. The number of transmission steps increases with the volume, because of the higher density of small eigenvalues, so we expect the transmission/refection part of the algorithm to dominate. The auto-correlation time should decrease as the number of successful topological index changes increases. An efficient algorithm will therefore have as few reflections as possible, and therefore a maximised transmission rate. 

In this paper we demonstrate that by using these methods we can improve the tunnelling of the overlap Hybrid Monte Carlo by a significant factor, and the cost of the factorisation is far less than the gain from reduced auto-correlation. This gain will increase at smaller mass, while the cost does not. Using this method, even at small mass, the transmission/reflection part of the algorithm will therefore allow for frequent
 topological tunnelling.  

This article is organised as follows: in section \ref{sec:1b}, we present a model that approximates the scaling of the action jump with the quark mass for both the actual determinant and the pseudo-fermion estimate. Section \ref{sec:2} describes our proposed factorisation, and the (standard) method which we will compare it against in our tests. Section \ref{sec:3} describes our numerical results, and, after a few comments on energy conservation and the fermionic force in section \ref{sec:3b}, we conclude in section \ref{sec:4}.

\section{The scaling of the action jump with the quark mass.}\label{sec:1b}
In subsequent sections, we will present a method which vastly improves the scaling of the action jump with the quark mass. To motivate our methods and interpret our results we have developed a simple model to estimate the scaling of the action jump when estimated by pseudo-fermions and for the actual fermion determinant. We stress that while this somewhat simple model shows properties which we observe numerically, our algorithm does not depend on this model in any way. 

The discontinuity in the action for the pseudo-fermion discontinuity is ~\cite{Cundy:2005pi}
\begin{align}
\Delta S =& \bra{\phi}\frac{1}{H_{+}^2}(H_-^2 - H_+^2)\frac{1}{H_{-}^2}\ket{\phi}\nonumber\\
=&(1-\mu^2)\bra{\phi}\frac{1}{H_{+}^2} (\gamma_5\ket{\psi}\bra{\psi} + \ket{\psi}\bra{\psi}\gamma_5)\frac{1}{H_{-}^2}\ket{\phi}2\epsilon(\lambda_-),\label{eq:ds}
\end{align}
where $\psi$ is the Kernel eigenvector whose eigenvalue is zero, and the $-$ subscript indicates the overlap operator just before the crossing and the $+$ subscript just after the crossing, and 
\begin{gather}
\ket{\phi} = H_0 \ket{\eta},
\end{gather}
where $H_0$ is the overlap operator at the start of the trajectory. If the eigenvectors of $\gamma_5 H $ are $\ket{\theta}$, with eigenvalues $1+\mu + (1-\mu)e^{i\theta}$, then it is easy to show that $H\ket{\theta} = (1+\mu + (1-\mu)e^{i\theta})\gamma_5\ket{\theta}$, and that $H^2\ket{\theta} = (2+ 2\mu^2 + (1-\mu^2)(e^{i\theta}+e^{-i\theta}))\ket{\theta}$. 
Inserting the complete set of the overlap eigenvectors into equation (\ref{eq:ds}) gives
\begin{align}
\Delta S =& 2\epsilon(\lambda_-)\bra{\eta}\sum_{\theta_0}\sum_{\theta_0'}\sum_{\theta_-}\sum_{\theta_+} \gamma_5\ket{\theta_0}\braket{\theta_0}{\theta_+}\bra{\theta_+}\gamma_5\ket{\psi}\braket{\psi}{\theta_-}\braket{\theta_-}{\theta'_0}\bra{\theta'_0}\gamma_5\ket{\eta}\nonumber\\
& \frac{1+\mu + (1-\mu)e^{i\theta_0}}{2+2\mu^2 + (1-\mu^2)(e^{i\theta_+} + e^{-i\theta_+})}\frac{{1+\mu + (1-\mu)e^{-i\theta'_0}}}{2+2\mu^2 + (1-\mu^2)(e^{i\theta_-} + e^{-i\theta_-})} + h.c..\label{eq:ds2}
\end{align}
To proceed further, in our efforts to calculate $\langle \Delta S \rangle$, we need to make a number of simplifying assumptions. Firstly, we need to simplify the matrix elements obtained from the inner products of the eigenvectors. We assume, based on the normalisation of the vectors,  that we can treat terms such as $\braket{\eta}{\theta}$ as inversely proportional to the square root of the volume, and that the other matrix elements are inversely proportional to the volume. We also assume that we can average the scalar products so that they are analytic functions of the eigenvalues, so $\langle\braket{\psi}{\theta}\rangle = \psi(\theta)$, $\langle\bra{\psi}\gamma_5\ket{\theta}\rangle = \psi_5(\theta)$, $\langle\braket{\theta_1}{\theta_2}\rangle = \chi(\theta_1)\chi(\theta_2)$, and $\langle\bra{\eta}\gamma_5\ket{\theta}\rangle = \eta_5(\theta)$. This incorporates the assumption that all the matrix elements are statistically independent. In particular, this means that the molecular dynamics has progressed sufficiently far that there is no correlation between $\ket{\theta_0}$ and $\ket{\theta_{\pm}}$. While this might seem improbable, we stress that we make this assumption only to simplify the presentation. The argument below, at a qualitative level, only requires that the average value of the product of the matrix elements is an analytic function of the eigenvalues. Secondly, we assume that we can neglect the zero modes of the overlap operator, since the density of zero modes scales as $\sqrt{V}$, where $V$ is the lattice volume, while elements of the matrix $\ket{\chi_-}\bra{\chi_-}$ scale inversely with the volume. We also assume that the density of overlap eigenvalues, $\rho(\theta)$, is a real and analytic function in the complex plane, that it has a well defined non-infinite value on the circle of infinite radius and that it is also an analytic function of the quark mass.\footnote{Attempts to construct the eigenvalue density from the fermion weight, $\prod_i \sum_i(2+2\mu^2 + 2(1-\mu^2)\cos\theta_i)$, and the gauge action, which, following ~\cite{Liu:2007hq} can be written as $e^{-\beta' \sum_i (2+2\mu^2 + 2(1-\mu^2)\cos\theta_i)}$ fail because of problems constructing the Jacobian $|\partial U/\partial \theta_i|$. Assuming that this Jacobian can be treated as a constant and that there is no correlation between the eigenvalues, following ~\cite{Golterman:2007ni}, fails in the p-regime because it leads to a vanishing density of zero eigenvalues at zero mass, which would lead to a zero chiral condensate through the Banks-Casher relation~\cite{Banks-Casher}.} The assumption of analyticity with respect to the quark mass should hold in large enough lattice volumes, but probably not, for example, in the transition between the $p$- and $\epsilon$-regimes (if this is a phase-transition rather than a crossover), which are known to have different eigenvalue densities. So, we could expect a significantly different action jump between these different regimes, although the action jump within the various regimes of QCD should vary smoothly with the mass.
 
 Clearly these are severe assumptions (without which is is difficult to see how to proceed), so results from this model should only be seen as tentative, and need to be confirmed numerically. Using these simplifications, we can now evaluate each term in equation (\ref{eq:ds2}) separately. We achieve this by substituting $z = e^{i\theta}$ and integrating along the unit circle in the complex plane. So,
\begin{align}
&\int_0^{2\pi} d\theta \rho(\theta)\eta_5(\theta)\chi(\theta) \left[(1+\mu) + e^{i\theta}(1-\mu)\right] =\phantom{a} \nonumber\\&
\oint \frac{dz}{iz}\rho(\theta)\eta_{5}(\theta)\chi(\theta)\left[1 + \mu + z (1-\mu)\right] =\nonumber\\
&\phantom{evenlotsmorelovelyspace}2\pi(1+\mu)\eta_5(-i\infty)\chi(-i\infty) \rho(-i\infty),\label{eq:rubbishmodel1}
\end{align}
and
\begin{align}
\int_0^{2\pi} d\theta& \rho(\theta)\chi(\theta)\psi(\theta) \frac{1}{1+\mu^2 + 2(1-\mu^2)\cos\theta}\nonumber\\=&\oint dz \frac{1-\mu^2}{4i\mu}\left[\frac{1}{(1-\mu^2)z + 1 + \mu^2 - 2\mu} - \right.\nonumber\\&\phantom{lots of really lovely space}\left. \frac{1}{(1-\mu^2)z + 1 + \mu^2 + 2\mu}\right]\rho(\theta)\chi(\theta)\psi(\theta)\nonumber\\
=&\frac{(1-\mu^2)\pi}{4\mu}\rho(-i\log(-\frac{(1-\mu)^2}{1-\mu^2}))\chi(-i\log(-\frac{(1-\mu)^2}{1-\mu^2}))\nonumber\\&\phantom{need a really really huge amount of space here} \psi(-i\log(-\frac{(1-\mu)^2}{1-\mu^2}).\label{eq:rubbishmodel2}
\end{align}
Inserting the results of equations (\ref{eq:rubbishmodel1}) and (\ref{eq:rubbishmodel2}) into (\ref{eq:ds2}) gives the dominant mass contribution to $\Delta S$ as O($\mu^{-2}$).

For the actual determinant, as opposed to the pseudo-fermion estimate, the action jump is ~\cite{DeGrand:2004nq,Egri:2005cx}
\begin{gather}
\Delta S = 2\log\left(1 + 2\epsilon(\lambda_-)\bra{\psi}\frac{1}{H_-}\ket{\psi}\right)\label{eq:realds}
\end{gather}
Using the same assumptions as before, we write
\begin{gather}
\bra{\psi}\frac{1}{H_-}\ket{\psi} = \sum_{\theta_-}\bra{\psi}\gamma_5\ket{\theta_-}\braket{\theta_-}{\psi}  \frac{1}{1+\mu + (1-\mu)e^{i\theta_-}}.\label{eq:16}
\end{gather}
Replacing the sum in equation (\ref{eq:16}) with an integral, and using the same contour as before, we note that the pole at $1+\mu + (1-\mu)z=0$ lies outside the contour and obtain
\begin{gather}
\left\langle\bra{\psi}\frac{1}{H_-}\ket{\psi}\right\rangle \propto \frac{2\pi}{1+\mu}\rho(-i\infty)
\end{gather}
Hence the discontinuity in the actual determinant when there is a topological charge change (and if our numerous assumptions are valid) does not diverge in the quark mass. By inspection of equation (\ref{eq:realds}), it can be shown that it is also independent of the lattice volume.

We outlined earlier that the probability of transmission when there is an attempted topological charge change is $\min(1,e^{-\Delta S})$, and numerical evidence indicates that, with a pseudo-fermion estimate of the determinant, $\Delta S$ is almost invariably positive.\footnote{This is suggested from considerations concerning reversibility. The number of transmission steps with positive $\Delta S$ should be equal to the number of transmission steps with negative $\Delta S$, because for a reversed trajectory $\Delta S \rightarrow -\Delta S$, and there should be no change in the number of topological charge changes. The number of topological charge changes is the number of attempted topological charge changes times the probability of transmission. The probability of transmission when $\Delta S$ is negative is one. Therefore the ratio of the number of attempted topological charge changes for positive $\Delta S$ to the number for negative $\Delta S$ is inversely proportional to the probability of a topological charge change when $\Delta S$ is positive. 
} The auto-correlation time for topological quantities, most obviously the susceptibility, clearly depends strongly on this transmission rate. For the sake of being definite, we will assume that the auto-correlation time is inversely proportional to the rate of topological charge changes, which itself is proportional to the transmission rate. Although this argument is somewhat over-simplified, numerical studies suggest that it is not unreasonable. Thus dynamical overlap simulations where the determinant is simulated with pseudo-fermions have an auto-correlation time that scales as $e^{-\alpha/\mu^2}$. Because all lattice Dirac operators have the same continuum limit, we expect similar behaviour for other formulations of lattice QCD, although, because these do not have a well defined topological index, the problem may not be as apparent as with overlap fermions.

One possible way of reducing the problem is to use multiple pseudo-fermions~\cite{DeGrand:2004nq,Hasenbusch:2001ne}. For example, simulating the fermion determinant using
\begin{gather}
\det H^2 = e^{\bra{\phi_1} \frac{1}{H(\mu_1)^2}\ket{\phi_1} + \bra{\phi_2} \frac{H(\mu_1)^2}{H(\mu)^2}\ket{\phi_2}}
\end{gather}  
gives
\begin{gather}
\Delta S = \alpha \mu_1^{-2} + \beta \frac{\mu_1}{\mu} + \gamma\frac{(\mu_1^2 - \mu^2)^2}{\mu^2} + O(\mu^0),
\end{gather}
where $\alpha$ and $\beta$ are constants which need to be determined.
In principle, by adding more pseudo-fermions, one can reduce the action discontinuity considerably. However, as the mass decreases, more pseudo-fermions need to be added, increasing the size of the fermionic force (once the optimum number of pseudo-fermions has been passed), the action jump, and the time required for each trajectory. Therefore this method, while useful, cannot produce a $\Delta S$ that is independent of the quark mass.

An alternative approach is to factorise the discontinuity from the pseudo-fermion term, and treat the discontinuous part of the action exactly. An obvious way of doing this (although we do not use this method) is to use a projector $P$ constructed from the smallest $n$ kernel eigenvectors. We can write 
\begin{align}
\det[H] =& \det\left[1-P\frac{1}{PHP} PH(1-P)\right]^{-1} \det\left[1-(1-P)HP\frac{1}{PHP}P\right]^{-1}\nonumber\\
&\det\left[PHP + (1-P)H(1-P) - (1-P)HP\frac{1}{PHP}PH(1-P)\right]\label{eq:det0}
\end{align}
If the projectors are exact, so that $P^2 = P$, then the first two determinants on the RHS of equation (\ref{eq:det0}) are 1 and the third determinant can be further factorised into the product of
\begin{gather}
\det[PHP + (1-P)]\label{eq:det1}
\end{gather}
and
\begin{gather}
\det[(1-P)H(1-P) - (1-P)HP\frac{1}{PHP}PH(1-P) + P].\label{eq:det2}
\end{gather}
This is, of course, the familiar Schur decomposition. The projectors $P$ could in principle be constructed using $P = 1-\sum_{i=0}^n \ket{\psi_i}\bra{\psi_i}$. The first determinant, (\ref{eq:det1}), is then continuous when a kernel eigenvalue crosses zero, and can be treated with pseudo-fermions. The second determinant, given in equation (\ref{eq:det2}), is discontinuous, but the matrix is small for the determinant to be calculated exactly, and explicitly included in the action without pseudo-fermions. Of course, the determinant given in equation (\ref{eq:det1}) is discontinuous when there is a level crossing between the $n$th and $(n+1)$th eigenvalues. To avoid this, without resorting to another transmission/reflection step,\footnote{Ergodicity could here be preserved by constructing the projector from a different number of eigenvalues on different trajectories; as long as care is taken so that this does not violate reversibility.} it is necessary to use projectors which are a function of the eigenvalue, which means that it is impossible to satisfy $P^2 = P$. While this decomposition is still possible, we cannot neglect the first two determinants in equation (\ref{eq:det0}), and the third determinant cannot be factorised so easily. For this reason, we prefer to use the simpler means to achieve the same end described in the next section (algorithm C).
\section{Determinant factorisation}\label{sec:2}
\subsection{Algorithm A}\label{sec:2a}
Algorithm A was introduced in the context of dynamical overlap simulations by DeGrand and Schaefer in~\cite{DeGrand:2004nq}. It reduces the noise of the pseudo-fermion approximation by introducing $n$ additional pseudo-fermions, following the method of Hasenbusch~\cite{Hasenbusch:2001ne}:
\begin{align}
\det(H(\mu)) =& \det\frac{H(\mu)}{H(\mu_1)}\det\frac{H(\mu_1)}{H(\mu_2)}\ldots \det(H(\mu_n))\\
S_{pf} = & \phi_1^{\dagger}H(\mu_1)H(\mu)^{-2}H(\mu_1)\phi_1+\phi_2^{\dagger}H(\mu_2)H(\mu_1)^{-2}H(\mu_2)\phi_2+\ldots +\phantom{a}\nonumber\\& \phi_{n+1}^{\dagger}H(\mu_n)^{-2}\phi_{n+1}
\end{align} 
This decomposition was originally introduced to precondition the fermionic force, and as such a few additional pseudo-fermions are needed for an efficient algorithm at low mass. As discussed in the previous section, it can also be used to effectively reduce the action discontinuity. We use algorithm A with one additional pseudo-fermion, as a benchmark against which we will compare the results of our new method. 
\subsection{Algorithm C}\label{sec:2c}
 We factorise the overlap determinant using an eigenvalue projector which depends on a continuous function $\alpha$ of the kernel eigenvalue.
\begin{align}
\tilde{H}_C = & (1+\mu)\gamma_5 + (1-\mu)\epsilon(Q)\left(1-\sum_i\alpha(\lambda_i)\ket{\psi_i}\bra{\psi_i}\right)\\
\det{H} =& \det[\tilde{H}_C]\det\left[ 1 + (1-\mu)(\epsilon(Q)\sum_i \alpha(\lambda_i)\ket{\psi_i}\bra{\psi_i} )\frac{1}{\tilde{H}_C}\right],\label{eq:26}
\end{align} 
where  $\alpha(0) =1$ and $\alpha(\lambda \ge \Lambda) = 0$ for a carefully tuned eigenvalue cutoff $\Lambda$. 
The first of the determinants in equation (\ref{eq:26}) is a continuous function, so it will not contribute to the action jump. The second can be written as
\begin{align}
D_C =& \det\left[1 + (1-\mu)\ket{\psi_i}\bra{\psi_i}\left(\epsilon(\lambda_i)\alpha(\lambda_i)\right)\frac{1}{\tilde{H}_C}\right]\nonumber\\
=&\det\left[\delta_{ij} + (1-\mu)\bra{\psi_i}\left(\epsilon(\lambda_i)\alpha(\lambda_i)\right)\frac{1}{\tilde{H}_C}\ket{\psi_j}\right].
\end{align}
This is the determinant of an $n\times n$ matrix, which can be calculated easily using, for example, LU decomposition~\cite{Numericalrecipies}. It has been proposed to account for this determinant by re-weighting, or by including the correct expression in the Hybrid Monte-Carlo Metropolis step, but this leads to a low acceptance rate and hence an algorithm that is only a small improvement over algorithm A~\cite{Schaefer:2006bk}. We propose adding the logarithm of this determinant to the fermion action. This obviously costs more per trajectory, both because of the additional term in the fermionic force, and because, since $D_C$ is discontinuous at the eigenvalue crossing, we would need to track the eigenvalues of the kernel operator and use the reflection/transmission routine to account for the discontinuity in the action.\footnote{It might be possible to remove some of this additional cost because, as outlined in section \ref{sec:1b} and demonstrated in section \ref{sec:3}, we expect this factorisation to reduce  $\Delta S$, and if it reduces it enough (to, for example, $\sim 0.5$), then it might be possible to neglect the transmission/reflection step, accepting that the algorithm will not conserve energy, using the final HMC metropolis step to ensure detailed balance.  This would lead to a lower HMC acceptance rate, but the time per trajectory would be faster: whether neglecting transmission/reflection would be efficient will depend on the size and spread of $\Delta S$ and the frequency of attempted topological charge changes.} However, including the additional term in the action gains by maintaining a high Monte-Carlo acceptance rate, and we believe that this gain outweighs the losses, at least on moderate size lattices. In our numerical tests, we used $\alpha(\lambda) = \epsilon(\lambda) Z(\lambda)$, where $Z$ is the same Zolotarev rational approximation we use to approximate the sign function~\cite{vandenEshof:2002ms}. With this choice, there is no need to project out the eigenvalues when calculating $\tilde{H}_C$. In principle, the choice of $\alpha$ is arbitrary, although we should seek to maximise the HMC acceptance rate and avoid exceptional configurations.

The fermion action is thus (with no additional Hasenbusch pseudo-fermion terms):
\begin{gather}
S_{pf} = \phi^{\dagger}\frac{1}{\tilde{H}_C}\phi - 2\log(D_C)
\end{gather}
The determinant can be differentiated easily, by differentiating each component of the matrix, and multiplying it by the co-factor of that component. As an illustration, for a $3\times 3$ matrix, we write
\begin{align}
\frac{d}{d\tau}&\log(\det[a]) =\nonumber\\& \frac{1}{\det[a]}\frac{da_{11}}{d\tau}\det\left[\begin{array}{c c c}
1&0&0\\
0&a_{22}&a_{23}\\
0&a_{32}&a_{33}
\end{array}\right] + \frac{1}{\det[a]}\frac{da_{12}}{d\tau}\det\left[\begin{array}{c c c}
0&1&0\\
a_{21}&0&a_{23}\\
a_{31}&0&a_{33}
\end{array}\right]+\ldots.
\end{align}
We differentiate the eigenvectors and eigenvalues following established methods ~\cite{Cundy:2005pi,Cundy:2007df}.

\section{Numerical results}\label{sec:3}
We tested algorithms A and C on a $8^316$ lattice at masses $\mu = 0.03$, $\mu = 0.04$ and $\mu = 0.05$, corresponding to pion masses of $450-550~\text{MeV}$ (calculated from the pseudo-scalar correlator) at lattice spacings of about  $a = 0.11~\text{fm}$ (calculated using $r_0 = 0.49$~\cite{Sommer:1993ce}). These masses were as low as we could achieve on these lattice volumes while remaining in the topologically active $p$-regime. We used two steps of stout smearing with an otherwise standard Wilson kernel at smearing parameter $\rho = 0.1$, and for these simulations we used no  Sexton-Weingarten integration~\cite{SW} (which should not affect the size of the action jump). We used the Omelyan integrator~\cite{Omelyan}. We used one additional flavour of Hasenbusch fermions for both algorithms A and C, with the the Hasenbusch mass, $\mu_1$ the same for algorithms A and C, but varying according to quark mass in an (admittedly very crude) attempt to optimise $\Delta S$ for algorithm A. In fairness, we want to stress that our choice of $\mu_1$  turned out to be sub-optimal at $\mu = 0.04$. We generated between 120 and 500 trajectories for each of our ensembles, and all our ensembles ($\mu = 0.04$, algorithm C was the smallest in each case) had over 170 attempted topological index changes. We ran trajectories of length 0.5 with 20 integration steps, which achieved an Monte Carlo acceptance rate of over 90\% in all cases.

\begin{figure}
\begin{center}
\includegraphics{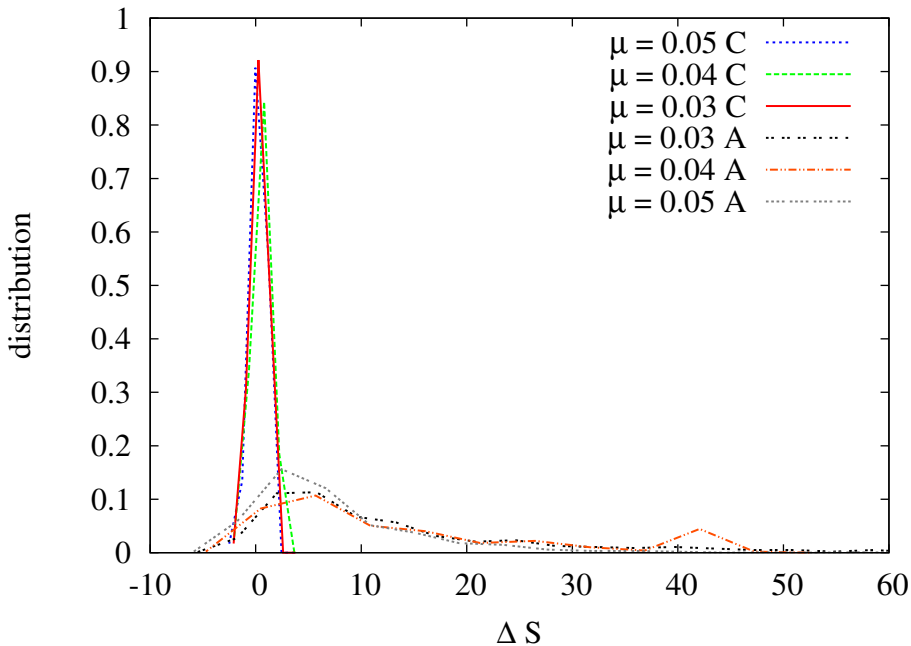}
\end{center}
\caption{The distribution of $\Delta S$ for masses $\mu = 0.03$, $\mu = 0.04$, and $\mu = 0.05$ for algorithms A and C.}\label{fig:2}
\end{figure}

 The mean values of the action jump at the topological sector boundary, $\Delta S$, standard deviation of $\Delta S$, and the transmission rates are given in table \ref{tab:1}, together with the Hasenbusch mass $\mu_1$. The distribution of $\Delta S$ is shown in figure 
\ref{fig:2}.

\begin{table}
\begin{center}
\begin{tabular}{l l|| l l||l l || l l}
$\mu$&$\mu_1$&A&C & A &C &A &C\\
\hline\hline
0.03&0.17&14.0(7)&0.28(8)
&21.9&0.78
&9.6\% &77.5\%
\\
0.04&0.22&13.8(10)&0.59(11)
&19.5&0.77
&9.2\% &75.2\%
\\
0.05&0.22&7.8(4)&0.24(7)
&13.6&0.72
&18.3\% &70.1\%
\end{tabular}
\end{center}
\caption{Average values of $\Delta S$ (left) standard deviations of $\Delta S$ (middle) and transmission rates (right) for algorithms A and C.}\label{tab:1} 
\end{table}

It can be seen that algorithm C has a far lower action jump than algorithm A, and a much smaller spread of $\Delta S$. This corresponds to a much larger transmission rate. We note that the pion masses used are still large, and we expect a far larger gain at more realistic masses. As expected, the mass dependence of the action jump for algorithm C is not significant, while for algorithm A it has a strong dependence on the quark mass.

The increase in the transmission rate corresponds to a reduction in the topological auto-correlation time of about a factor of 8. The extra cost of the additional inversions meant that it took twice as long to generate the trajectory (measured on the $\mu = 0.04$ ensemble), meaning that the overall gain of this method was roughly a factor of 4. This was without using any Sexton-Weingarten methods to place the small determinant on a reduced time-scale, which we believe will be effective, and at a relatively large quark mass. We expect larger gains for simulations at larger lattice volume and smaller mass.

\section{Volume scaling and energy conservation}\label{sec:3b}
Another attempt along similar lines has run into difficulties with energy conservation~\cite{Degrandschaeferpersonalcommunication}. Part of the problem encountered in these other studies could have been avoided by using the improved eigenvector differentiation of~\cite{Cundy:2007df}. Still, part of the problem could be caused by other issues which affect the size and stability of the fermionic force. It is therefore appropriate to make a few remarks about energy conservation, and in particular the performance on larger volumes.

The obvious disadvantage of our method is the need to perform numerous additional inversions of the overlap operator in the calculation of $D_C$. The number of extra inversions will depend on the density of kernel eigenvalues, and the value chosen for the cut-off $\Lambda$. In this section we will assume that to maintain a fixed number of small eigenvalues below the cut-off, we would have to reduce $\Lambda$ inversely proportional to the lattice volume, which seems to us to be the worst case scenario, as well as being suggested by the Banks-Casher relation~\cite{Banks-Casher} (we do not yet have any data to study this question). An option would be to increase the number of additional inversions as the volume increases. In practice, we have combined the two approaches so that we reduce $\Lambda$ but not to such an extent that the number of inversions in the small matrix remains constant. In this section, we will investigate the effect of reducing $\Lambda$  on the fermionic force and energy conservation. 


For the leapfrog QPQ integration scheme (where the gauge field is updated by a half step, the momentum by a full step, and the gauge field by another half step), there is a shadow Hamiltonian, $H'$, which is exactly conserved by the molecular dynamics for a trajectory of fixed length~\cite{Clark:2007ffa}
\begin{align}
H' =& H + \frac{1}{24}\left(-p^2S'' +2 {S'}^2\right)\delta\tau^2 +\phantom{a}\nonumber\\&\phantom{lots of space}\frac{1}{5760}\left(7p^4S^{(4)} - 24p^2(S'S''' + 2S''^2) +96{S'}^2S''\right)\delta\tau^4 + \ldots, 
\end{align}
where $S$ is the action, $p$ represents the molecular dynamics momentum, and the prime indicates its derivative with respect to the gauge field. From this expression, it is easy to read off the size of the energy conservation violation. In our case it is clear that $d\alpha/d\lambda$ is proportional to $1/\Lambda$, $d^2\alpha/d\lambda^2$ is proportional to $1/\Lambda^2$ and so on, and if $\alpha$ is continuous, we could continue this series indefinitely (bearing in mind that $\alpha$ and all its derivatives are zero, or close to zero, for $\lambda > \Lambda$). Thus ${S'}$ contains terms of order $1/\Lambda$, $S''$ terms of order $1/\Lambda^2$ and $1/\Lambda$, and so on. It is clear that there is a problem with the volume scaling, particularly from the higher order derivatives. We use the Omelyan integrator,~\cite{Omelyan,Takaishi:2005tz},
\begin{gather}
U' = e^{\alpha\tau \Pi \frac{\partial}{\partial q}}e^{-\frac{1}{2}\tau S'\frac{\partial}{\partial p}} e^{(1-2\alpha)\tau \Pi \frac{\partial}{\partial q}}e^{-\frac{1}{2}\tau S'\frac{\partial}{\partial p}}e^{\alpha\tau \Pi \frac{\partial}{\partial q}}U,
\end{gather}
which conserves the shadow Hamiltonian
\begin{gather}
H' = H + \delta\tau^2\left(\frac{6\alpha^2-6\alpha +1}{12}p^2S'' - \frac{1-6\alpha}{24}{S'}^2\right) + O(\delta\tau^4).
\end{gather}
By tuning $\alpha$ to $\frac{1}{2} - \frac{1}{\sqrt{12}}$ we are able to remove the troublesome ${S''}$ terms from the leading order correction to the shadow Hamiltonian. Obviously, this is not a complete solution to the problem, and we are in the process of developing a method which will allow the factorisation used here to be extended to larger volumes~\cite{cundyforthcoming}. However, this method is more than sufficient for our current runs on $8^316$ and $12^348$ lattices, where we achieve close to 100\% acceptance in both cases.

Obviously, another means to reduce this problem would be to reduce the density of small kernel eigenvalues, either through additional smearing or the tunnelling HMC algorithm of ~\cite{Golterman:2007ni}. We also note that deflation methods can be usefully employed when there are multiple right hand sides to solve. By using the eigenvectors from the previous molecular dynamics step as a initial guess, we are able to calculate the eigenvectors of $\tilde{H}_C$ to a suitable precision with only a few calls to a low accuracy sign function on our $8^316$ lattices. Unlike deflation methods with the overlap operator itself, there is no large shift in the eigenvectors of $\tilde{H}_C$ when there is a topological charge change, meaning that the calculation of the overlap eigenvalues consumes only a small fraction of the total computer time per trajectory.

\section{Conclusion}\label{sec:4}
The topological auto-correlation time limits dynamical lattice QCD simulations, particularly those involving overlap fermions. We have demonstrated that by factorising the fermion determinant into a large continuous matrix approximated by pseudo-fermions and a smaller discontinuous part treated exactly, we are able to observe topological tunnelling at all masses. With our new method, the discontinuity in the pseudo-fermion action is independent of the quark mass, and we expect it to be independent of the lattice volume.  There may be problems affecting the energy conservation and HMC acceptance rate on larger volumes, but we expect that these can be resolved.

\section*{Acknowledgements}
Calculations were performed on the JUBL and Cray XD-1 computers at The J\"ulich Supercomputing Center (JSC), at the Forschentrum J\"ulich, Germany. NC is grateful for the support of grant 930183 from the EU RP-6 "Hadron Physics" project, from the DFG "Gitter-Hadronen Ph\"anomenologie" project, number 458/14-4 and EU grant MC-EIF-CT-2004-506078.

\bibliographystyle{modified_cpc}
\bibliography{tunnelingpaper}

\end{document}